\documentclass[submission,copyright]{eptcs}
\providecommand{\event}{HSB 2013} 

\usepackage{graphicx}
\usepackage{amsmath}

\title{Producing a Set of Models for the Iron Homeostasis Network}

\author{Nicolas Mobilia
\institute{UJF-Grenoble 1 / CNRS\\TIMC-IMAG UMR 5525,\\Grenoble, F-38041, France}
\email{nicolas.mobilia@imag.fr}
\and
Alexandre Donz{\'e}
\institute{EECS Department\\University of California Berkeley\\Berkeley, CA 94720 USA}
\email{donze@eecs.berkeley.edu}
\and
Jean Marc Moulis
\institute{Univ. Grenoble Alpes, Laboratory of Fundamental and Applied Bioenergetics (LBFA),\\and Environmental and Systems Biology (BEeSy), Grenoble, France\\Inserm, U1055, Grenoble, France\\CEA-iRTSV, Grenoble, France}
\email{jean-marc.moulis@cea.fr}
\and
{\'E}ric Fanchon
\institute{UJF-Grenoble 1 / CNRS\\TIMC-IMAG UMR 5525,\\Grenoble, F-38041, France}
\email{eric.fanchon@imag.fr}
}

\def\titlerunning{A Model of the Cellular Iron Homeostasis Network}
\def\authorrunning{N. Mobilia, A. Donz{\'e}, J.M. Moulis, {\'E}. Fanchon}

\renewcommand{\d}{\mathrm{d}\,}
\newcommand{\TfR} {TfR1}
\newcommand{\TfRf}{TfR1_{_-f}}
\newcommand{\TfRb}{TfR1_{_-b}}
\newcommand{\TfRp}{TfR1_{_-p}}

\newcommand{\Ft} {Ft}
\newcommand{\Ftf}{Ft_{_-f}}
\newcommand{\Ftb}{Ft_{_-b}}
\newcommand{\Ftp}{Ft_{_-p}}

\newcommand{\FPNa} {\mathit{FPN1a}}
\newcommand{\FPNaf}{\mathit{FPN1a_{_-f}}}
\newcommand{\FPNab}{\mathit{FPN1a_{_-b}}}

\newcommand{\IRP}{IRP}
\newcommand{\Fe}{Fe}

\begin{document}

\maketitle

\section{Introduction}
\label{sec:introduction}

The continuous parts of complex biological systems are often modeled
by use of Ordinary Differential Equations (ODE). When experimental
data are available, they usually have large variability, for example
due to variability in cell cultures. Since data on a given system are
scarce, one generally uses data from different cell types, different
organisms, or different conditions. All this translates into large
parameter uncertainty. To cope with this situation and try to
integrate all available data in a consistent model, we represent such
data by intervals rather than single numerical values. The ranges of
the intervals vary depending on the type of experiment and the nature
of the experimental system. Some parameters or concentrations are not
known at all and are initially defined to belong to the physiological
domain. This set of intervals define the search space. Other
experimental data are expressed in terms of inequalities involving
derived quantities. Our goal is to build a set of models that satisfy
all the constraints deduced from experiments, and to analyze the
salient features of the dynamics of this set of models.

We present a method for modeling biological systems which combines
formal techniques on intervals, numerical simulations and formal
verification of STL (Signal Temporal Logic) formula. This allows us to
consider intervals for each parameter and to describe the expected
behavior of the model. We apply this method to the modeling of the
cellular iron homeostasis network in precursors of erythroid cells. A
core model~\cite{mobilia_donze_2012} has been presented
previously. Herein, we describe a more evolved model in which the
regulation mechanism acting at the translational level is explicitly
considered. This leads to a larger model with more parameters and the
integration of newly obtained experimental data. This new model
provides a more detailed description of the regulatory mechanism,
including quantitative considerations pertaining to the involved
species, and it should allow us to more precisely address pending
biological questions. The higher level of complexity of this model,
compared to the core model, required the development of a method to
characterize efficiently steady states.

In Section~\ref{sec:biological_system}, we describe the iron
homeostasis network, and, in Section~\ref{sec:model}, the
corresponding model. Then, in the Section~\ref{sec:method}, we
describe the method used. We finally explain the work that remains to
be done and conclude.

\section{Biological system}
\label{sec:biological_system}

\begin{figure}[!ht]
\centering
\includegraphics[width=.7\linewidth]{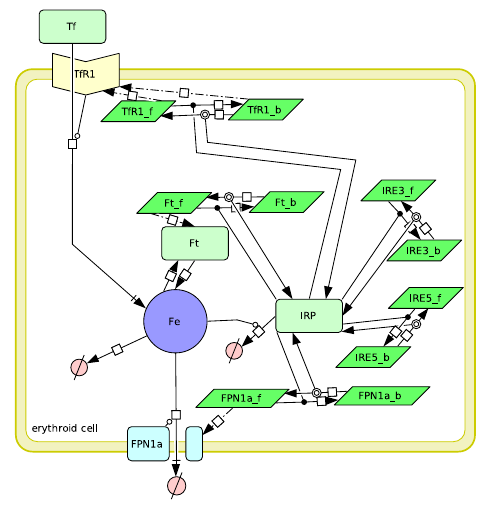}
\caption{Schematic representation of the main biological processes
  involved in the cellular control of iron concentration. This diagram
  was drawn with the software
  CellDesigner~\cite{funahashi_morohashi_2003}. The dashed arrows
  represent translation of mRNA into proteins. The lines ending with a
  combined perpendicular stroke and arrow represent iron transport
  through membranes. The regular arrows leading to an empty set symbol
  which indicate either degradation (for $\IRP$) or internal
  consumption (for iron). Moreover, the multi-arrows containing a
  black dot represent complexation while the ones with two empty
  circles mean decomplexation. Finally, the two regular arrows
  represent the loading/unloading of iron into/from the ferritins. The
  rounded rectangles represent proteins, the parallelograms represent
  mRNA, and the circle labeled $\Fe$ represents the pool of available
  iron. The concave hexagon represents the transferrin receptor. The
  species $IRE5_{_-f}$ (resp. $IRE3_{_-f}$) and $IRE5_{_-b}$
  (resp. $IRE3_{_-b}$) represent all the mRNA having an IRP binding
  site (called IRE) in the \mbox{5'-UTR} (resp. \mbox{3'-UTR)}
  excepted the ones explicitly drawn.\label{fig:model_graph}}
\end{figure}

Iron is an essential element for mammalian cells (eg. hemoglobin
contains iron), but if present in too high quantity iron has a
deleterious effect. The level of available iron is thus finely tuned
in mammalian cells. Our goal is to describe and understand this
regulatory mechanism. The regulatory network, described in
Figure~\ref{fig:model_graph}, is composed of fifteen species. The
species $\Fe$ (pool of available iron), $\IRP$ (Iron Regulatory
Protein), $\Ft$ (ferritin), $\FPNa$ (ferroportin) and $\TfR$
(transferrin receptor) were present in the previous
model~\cite{mobilia_donze_2012}. The other ones are the mRNA of these
proteins either in a free form or in a form complexed with an IRP. The
central regulatory mechanism, based on the IRP is described in
Mobilia~\&~al~\cite{mobilia_donze_2012}.

In a nutshell, the available data belong to several categories. A
qualitative description of the dynamics, obtained from a large body of
biological experiments is the following: if the amount of iron is
sufficient, the cell is in a steady state. From this state, if an iron
input cut-off occurs, the amount of iron in cells decreases and the
IRP are activated, leading to increased binding of IRP to
IRE-containing mRNAs. New kinetic parameters have been measured, as
well as absolute mRNA concentrations in the iron-replete regime. Our
aim is to build models which simulate the behavior of the biological
system. From the iron-replete steady state, the evolution of the
concentrations of the different species must be qualitatively
reproduced upon cutting off the iron supply.

\section{Model}
\label{sec:model}

We model this system with fifteen differential equations. These
equations contain 28 parameters. In the following sub-section, we
exhibit the equations related to the ferritin, the transferrin
receptor and the IRP. The conventions for the parameter names are the
following: a parameter named $p_{_-X}$ represents the transcription
speed of the mRNA of $X$ (units: mol/(L$\cdot$s)); $t_{_-X}$
represents the reaction rate constant for $X$ mRNA translation (units:
s$^{-1}$) ; $dp_{_-X}$ represents the degradation rate of the protein
$X$ (units: s$^{-1}$); $dr_{_-X}$ represents the degradation rate of
the mRNA of $X$ (units: s$^{-1}$); $drs_{_-X}$ represents the
degradation rate of the mRNA of $X$, when this mRNA contains an IRE in
the 3'-UTR region and is bound to an IRP (units: s$^{-1}$). The
species ending with the subscript $_{_-p}$ represent proteins, while
the ones ending with the subscript $_{_-f}$ (resp. $_{_-b}$) represent
free (resp. bound) mRNA concentration (units: mol/L).

\subsection{Ferritin equations}
\label{sec:ferritin_equations}

The ferritin mRNAs contain an IRE in the 5'-UTR region, so the
translation speed is proportional to the free mRNA concentration. This
is described by the first term in Equation~(\ref{eq:Ftp}). Moreover,
as a ferritin is constituted by 24 sub-units, a factor 1/24 appears in
this term. The second term represents the spontaneous degradation of
the ferritin.
\begin{equation}
\dfrac{\d\Ftp}{\d t} = \left(t_{_-Ft}/24\right)\cdot\Ftf - dp_{_-Ft}\cdot\Ftp
\label{eq:Ftp}
\end{equation}

The free ferritin mRNA concentration, described in
Equation~(\ref{eq:Ftf}), depends on four terms. The first one is the
transcription speed $p_{_-Ft}$. The second one represents the
complexation of this mRNA with an IRP (the parameter $ka$ is the
complexation second-order reaction rate parameter), while the third
one represents the decomplexation of mRNA:IRP complex. The parameter
$K_d$ is equal to the ratio $kd/ka$, where $kd$ is the decomplexation
first-order reaction rate parameter. The last term represents the
spontaneous degradation of the mRNA.
\begin{equation}
\dfrac{\d\Ftf}{\d t} = p_{_-Ft} - ka\cdot\Ftf\cdot\IRP + ka\cdot K_d\cdot\Ftb - dr_{_-Ft}\cdot\Ftf
\label{eq:Ftf}
\end{equation}

Finally, the bound ferritin mRNA concentration is described in
Equation~(\ref{eq:Ftb}). This equation is composed of three terms. The
first two, describing the complexation and the decomplexation, have
the same meaning as in the equation of the free ferritin mRNA. The
last term describes the spontaneous degradation of the mRNA.
\begin{equation}
\dfrac{\d\Ftb}{\d t} = ka\cdot\Ftf\cdot\IRP - ka\cdot K_d\cdot\Ftb - dr_{_-Ft}\cdot\Ftb
\label{eq:Ftb}
\end{equation}

\subsection{Transferrin receptor equation}
\label{sec:transferrin_receptor_equation}

The transferrin receptor mRNA contains five IREs in its 3'-UTR
region. Here, we make a simplification and consider that TfR1 mRNA
contains only one IRE. Equation~(\ref{eq:TfRp}) describes the
transferrin receptor concentration. It contains a translation term and
a spontaneous degradation term. As the IRE is located in the 3'-UTR
region, both free and bound mRNA are translated into
proteins. Moreover, because the receptor is a dimer, a factor 1/2
appears in the first term.
\begin{equation}
\dfrac{\d\TfRp}{\d t} = \left(t_{_-TfR1}/2\right)\cdot (\TfRf+\TfRb) - dp_{_-TfR1}\cdot\TfRp
\label{eq:TfRp}
\end{equation}

The equation of the free transferrin receptor mRNA is very similar to
the free ferritin mRNA. This equation is shown in
Equation~(\ref{eq:TfRf}) and is composed of a translation term, a
complexation term, a decomplexation term and a spontaneous degradation
term.
\begin{equation}
\dfrac{\d\TfRf}{\d t} = p_{_-TfR1} - ka\cdot\TfRf\cdot\IRP + ka\cdot K_d\cdot\TfRb - dr_{_-TfR1}\cdot\TfRf
\label{eq:TfRf}
\end{equation}

Equation~(\ref{eq:TfRb}) describes the concentration of the
transferrin receptor mRNA complexed with an IRP. This equation is
similar to the bound ferritin mRNA, except that the binding of an IRP
on this mRNA leads to the mRNA stabilization. To model this mechanism,
a specific degradation rate parameter ($drs_{_-TfR1}$) is
considered. This parameter is lower than the free mRNA degradation
rate parameter ($dr_{_-TfR1}$).
\begin{equation}
\dfrac{\d\TfRb}{\d t} = ka\cdot\TfRf\cdot\IRP - ka\cdot K_d\cdot\TfRb - drs_{_-TfR1}\cdot\TfRb
\label{eq:TfRb}
\end{equation}

\subsection{IRP equation}
\label{sec:irp_equation}

The IRP equation is described in Equation~(\ref{eq:IRP}).
\begin{eqnarray}
\dfrac{\d\IRP}{\d t}&=&{}- \left(\Ftf+\FPNaf+\TfRf+IRE3_{_-f}+IRE5_{_-f}\right)\cdot ka\cdot\IRP\label{eq:IRP}\\
&&{}+ \left(\Ftb+\FPNab+\TfRb+IRE3_{_-b}+IRE5_{_-b}\right)\cdot ka\cdot K_d\nonumber\\
&&{}- k_{Fe{\rightarrow}IRP}\cdot sig^+(\Fe,\theta_{Fe{\rightarrow}IRP})\cdot\IRP - dp_{_-IRP}\cdot\IRP\nonumber
\end{eqnarray}
The first line of this equation describes the complexation of free
mRNAs and IRP for all mRNAs, while the second line describes the
decomplexation of the bound mRNAs. The last line describes IRP
inactivation. This inactivation is described by a constant basal term
for degradation ($dp_{_-IRP}\cdot\IRP$) and the iron-triggered
regulation
($k_{Fe{\rightarrow}IRP}\cdot{}sig^+\left(\Fe,\theta_{Fe{\rightarrow}IRP}\right)\cdot\IRP$). Then,
if the iron level is significantly below the threshold
$\theta_{Fe{\rightarrow}IRP}$, the degradation rate is
$dp_{_-IRP}\cdot\IRP$. Otherwise, if the iron concentration is
significantly above this threshold, the degradation rate is
$(k_{Fe{\rightarrow}IRP}+dp_{_-IRP})\cdot\IRP$, where
$k_{Fe{\rightarrow}IRP}$ is the parameter describing the inhibition of
IRP by iron.

\subsection{Other equations}
\label{sec:other_equations}

The iron equation is the same than in the previous
model~\cite{mobilia_donze_2012}. The eight remaining equations are
very similar to those shown above. The equations for ferroportin and
$IRE5$ are similar to that of ferritin, and the equations for the
$IRE3$ species are similar to that of the transferrin receptor.

\subsection{Data}
\label{sec:data}

For each parameter, we consider an interval deduced from biological
data or incorporating meaningful values. For example, in K562 cells,
the ferritin half-life is 11 hours~\cite{kidane_sauble_2006}. We
deduce that the parameter $dp_{_-Ft}$ is included in the interval
\mbox{[3.8e-6, 3.8e-5]} s$^{-1}$.

Moreover, some data are expressed as relations between parameters. To
give an example, our recent data indicate the ferritin mRNA
concentration largely exceed that of the other IRP targets in
proliferating cells. The total ferritin (resp. IRE5) mRNA
concentration at steady state being equal to $p_{_-Ft}/dr_{_-Ft}$
(resp. $p_{_-IRE5}/dr_{_-IRE5}$), it follows the relation described in
Equation~(\ref{eq:rel_ft_ire5}).
\begin{equation}
\dfrac{p_{_-Ft}}{dr_{_-Ft}} > \dfrac{p_{_-IRE5}}{dr_{_-IRE5}}
\label{eq:rel_ft_ire5}
\end{equation}
We can also note that the lower degradation rate due to the binding of
IRP on IRE in 3'-UTR mRNA region translates into
Equation~(\ref{eq:rel_tfr1_deg}) and Equation~(\ref{eq:rel_ire3_deg}).
\begin{equation}
drs_{_-TfR1} < dr_{_-TfR1}
\label{eq:rel_tfr1_deg}
\end{equation}
\begin{equation}
drs_{_-IRE3} < dr_{_-IRE3}
\label{eq:rel_ire3_deg}
\end{equation}

In addition, some relations describe data related to the stationary
state. For example, the degradation rate of total TfR1 mRNA belong to
the interval \mbox{[7.0$\times10^{-6}$, 7.0$\times10^{-5}$]
  s$^{-1}$}~\cite{sharova_sharov_2009}\cite{seiser_posch_1995}. This
describe the sum of the degradation of both free and complexed mRNA
and translates into the equation~(\ref{eq:rel_tfr1}). The superscript
$^{eq}$ indicates that we consider the steady state concentration.
\begin{equation}
\dfrac{dr_{_-TfR1}\cdot\TfRf^{eq} + drs_{_-TfR1}\cdot\TfRb^{eq}}{\TfRb^{eq}+\TfRf^{eq}}\in\text{[}7.0\times10^{-6},~7.0\times10^{-5}\text{] s}^{-1}
\label{eq:rel_tfr1}
\end{equation}

The last kind of data is related to the dynamic of the system when an
iron cut-off happens. The modeling of these data using STL formula is
described in Mobilia~\&~al~\cite{mobilia_donze_2012}.

\section{Method}
\label{sec:method}

The set of intervals and constraints can be divided in two: those
pertaining to the iron-replete steady state, and those pertaining to
the cell response to iron shortage. In our previous
work~\cite{mobilia_donze_2012}, we first reduced the search space by
using the interval solver
Realpaver~\cite{granvilliers_benhamou_2006}. Then, we represented
formally the whole set of constraints as an STL formula and devised a
search algorithm to satisfy it, based on the tool
Breach~\cite{donze_2010}. Basically, a point is randomly drawn in the
search space, a simulation is performed and the STL formula is
evaluated. In the present more complete model, no iron-replete steady
state was initially found following the same procedure. In addition,
this failure did not instruct us on the origin of the problem.

To cope with this limitation, we improved the method in two
ways. (i)~In the first step, the interval solver Realpaver allowed to
reduce the intervals by propagating the constraints. Instead of being
solely a hyper-rectangle as previously, the search space was allowed
to be a union of hyper-rectangles thus reducing it more
efficiently. In case the interval solver found an inconsistency, we
improved the method by looking for the smallest sets of constraints
that have to be lifted to release the inconsistency. This information
gives insight when one wants to revisit the model and the data
used. (ii)~We decomposed the search algorithm in two parts. We had
developed an algorithm to generate efficiently a large number of
explicit solutions (steady states concentrations and model parameters)
satisfying the constraints of a stable steady state (the set of
constraint contains algebraic equations and inequalities involving
polynomial expressions). These explicit solutions, that were
prerequisite to perform simulation of the dynamics of the system, were
then fed into our Breach-based procedure in order to search models
satisfying the STL formula specifying the cell response to iron
deprivation.

The unknowns of the problem are the model parameters and the
concentrations in the iron-replete steady state. The methodology
proceeds basically as follows:
\begin{enumerate}
\item perform interval reduction with Realpaver;
\item select a subset of unknowns to be sampled (we start with
  unknowns within a narrow interval, then other criteria are used to
  decouple the equations and to optimize the following step);
\item for each sample of this subset of unknowns:
\item replace the instantiated unknowns in the algebraic equations and
  perform deductions (new unknowns can get instantiated);
\item check domain of newly instantiated unknowns;
\item check the validity of inequality constraints
  as soon as possible;
\item for each sub-problem (i.e.: set of decoupled constraints), apply
  this algorithm;
\item loop to step 3. until all samples are tried;
\item loop to step 2.
\end{enumerate}
The basic principles underlying this search are to decouple the
constraints in order to solve subproblems, and to identify the hardest
sub-problems (most constrained) and try to solve them in priority. The
aim is of course to trim the branches of the search tree as soon as
possible.

The interval solver Realpaver, used during the first step, allows to
reduce the intervals by propagating the constraints. The result is an
hyper-rectangle (or an union of hyper-rectangles) containing the
solutions, if they exist. The existence of solutions is not
guaranteed, but it is certain that there are no solutions outside of
the volume given by Realpaver. Consequently, this step is important to
reduce the search space. Nevertheless, the remaining space may be very
large with regard to the solution space. A simple example to
illustrate this aspect is the following: consider two unknowns $x_1$
and $x_2$, within the \mbox{[0, 1]} interval, and the constraint
$abs(x_1-x_2) < eps$, with $eps$ small compared to $x_1$ and $x_2$. If
the solution space is defined by one box, Realpaver cannot reduce the
search space. Considering a union of boxes allows a reduction of the
search space. As it is hard to find explicit solutions, we say
informally that the constraint is hard to satisfy (the smaller $eps$,
the harder it is).

The constraint system is constituted of algebraic equations and
inequalities. For the majority of them, the algebraic equations are
used to deduce the values of unknowns, and are thus automatically
satisfied. Inequalities are checked \textit{a posteriori}. To be
efficient, it is important to check an inequality as soon as all the
unknowns in it have been instantiated. For efficiency reasons,
redundant constraints are also added. The aim is to add constraints
simpler than the initial ones , which can be checked early in the
search process. Typically, from the constraints described by
Equation~(\ref{eq:rel_tfr1}), we straightforwardly deduce that
\mbox{$t_{_-TfR1}\cdot\TfRf^{eq}<5.5\times10^{-13}$} and that
\mbox{$t_{_-TfR1}\cdot\TfRb^{eq}<5.5\times10^{-13}$}. Even if $\TfRf$
or $\TfRb$ is not instantiated, one of these two constraints can be
checked and may invalidate this partial instantiation.

When applying this algorithm, we store the result of each verification
of domain and check of constraints (whether the constraint or domain
is verified or not). When no solution is found, this may be due to
different constraints. This information thus provide the level of
difficulty to satisfy each constraint. This may help to manually found
inconsistencies between constraints that were not automatically found
by Realpaver.

This methodology either provides us valid sets of values, or indicates
the hardest subset of constraints to satisfy. We applied it on the
iron homeostasis network. No solutions could be found, which is not a
proof of nonexistence, but the subset of hard constraints identified
allowed us to prove that there was indeed a contradiction. After
revision of this part of the model (namely: removing one non-reliable
constraint and extending some intervals), the procedure then generated
thousands of valid steady states in a short execution time.

\vspace{-4pt}
\section{Conclusion}
\label{sec:conclusion}

This evolved model describes in a more realistic way the action of IRP
and takes into account the fact that their effect depends on the
location of the binding site on mRNA. Moreover, it easily incorporates
new information obtained on the system.

Nevertheless, some work still remains to be done in order to
completely automate this search, and to interface the steady state
search with the part dealing with dynamical behavior (specified with
an STL formula).

Here we described our approach to nicely integrate different kinds of
biological data, combining an interval solver, simulations, and
temporal STL formula verification. We are applying it on a new model
of iron homeostasis in mammalian cells.

\vspace{-13pt}
\section*{Acknowledgements}
This work was supported by Microsoft Research through its PhD
Scholarship Programme and by the following grants: R{\'e}gion
Rh{\^o}ne-Alpes Cible 2010, IXXI-Spring 2012, and
Univ.~J.~Fourier-Agir 2013.

\vspace{-13pt}
\bibliographystyle{eptcs}
\bibliography{biblio_local}

\begin{thebibliography}{1}
\providecommand{\bibitemdeclare}[2]{}
\providecommand{\surnamestart}{}
\providecommand{\surnameend}{}
\providecommand{\urlprefix}{Available at }
\providecommand{\url}[1]{\texttt{#1}}
\providecommand{\href}[2]{\texttt{#2}}
\providecommand{\urlalt}[2]{\href{#1}{#2}}
\providecommand{\doi}[1]{doi:\urlalt{http://dx.doi.org/#1}{#1}}
\providecommand{\bibinfo}[2]{#2}

\bibitemdeclare{inproceedings}{donze_2010}
\bibitem{donze_2010}
\bibinfo{author}{A.~\surnamestart Donz{\'e}\surnameend} (\bibinfo{year}{2010}):
  \emph{\bibinfo{title}{Breach, A Toolbox for Verification and Parameter
  Synthesis of Hybrid Systems}}.
\newblock In: {\sl \bibinfo{booktitle}{CAV}}, pp. \bibinfo{pages}{167--170},
  \doi{10.1007/978-3-642-14295-6\_17}.

\bibitemdeclare{article}{funahashi_morohashi_2003}
\bibitem{funahashi_morohashi_2003}
\bibinfo{author}{A.~\surnamestart Funahashi\surnameend},
  \bibinfo{author}{M.~\surnamestart Morohashi\surnameend},
  \bibinfo{author}{H.~\surnamestart Kitano\surnameend} \&
  \bibinfo{author}{N.~\surnamestart Tanimura\surnameend}
  (\bibinfo{year}{2003}): \emph{\bibinfo{title}{CellDesigner: a process diagram
  editor for gene-regulatory and biochemical networks}}.
\newblock {\sl \bibinfo{journal}{BIOSILICO}}
  \bibinfo{volume}{1}(\bibinfo{number}{5}), pp. \bibinfo{pages}{159--162},
  \doi{10.1016/S1478-5382(03)02370-9}.

\bibitemdeclare{article}{granvilliers_benhamou_2006}
\bibitem{granvilliers_benhamou_2006}
\bibinfo{author}{L.~\surnamestart Granvilliers\surnameend} \&
  \bibinfo{author}{F.~\surnamestart Benhamou\surnameend}
  (\bibinfo{year}{2006}): \emph{\bibinfo{title}{{Algorithm 852}: {RealPaver}:
  an interval solver using constraint satisfaction techniques}}.
\newblock {\sl \bibinfo{journal}{{ACM} Transactions on Mathematical Software}}
  \bibinfo{volume}{32}(\bibinfo{number}{1}), pp. \bibinfo{pages}{138--156},
  \doi{10.1145/1132973.1132980}.

\bibitemdeclare{article}{kidane_sauble_2006}
\bibitem{kidane_sauble_2006}
\bibinfo{author}{T.~Z. \surnamestart Kidane\surnameend},
  \bibinfo{author}{E.~\surnamestart Sauble\surnameend} \&
  \bibinfo{author}{M.~C. \surnamestart Linder\surnameend}
  (\bibinfo{year}{2006}): \emph{\bibinfo{title}{{R}elease of iron from ferritin
  requires lysosomal activity}}.
\newblock {\sl \bibinfo{journal}{Am. J. Physiol., Cell Physiol.}}
  \bibinfo{volume}{291}(\bibinfo{number}{3}), pp. \bibinfo{pages}{C445--455},
  \doi{10.1152/ajpcell.00505.2005}.

\bibitemdeclare{article}{mobilia_donze_2012}
\bibitem{mobilia_donze_2012}
\bibinfo{author}{N.~\surnamestart Mobilia\surnameend},
  \bibinfo{author}{A.~\surnamestart Donz{\'e}\surnameend},
  \bibinfo{author}{J.~M. \surnamestart Moulis\surnameend} \&
  \bibinfo{author}{{\'E}.~\surnamestart Fanchon\surnameend}
  (\bibinfo{year}{2012}): \emph{\bibinfo{title}{A Model of the Cellular Iron
  Homeostasis Network Using Semi-Formal Methods for Parameter Space
  Exploration}}.
\newblock {\sl \bibinfo{journal}{EPTCS}} \bibinfo{volume}{92}, pp.
  \bibinfo{pages}{42--57}, \doi{10.4204/EPTCS.92.4}.

\bibitemdeclare{article}{seiser_posch_1995}
\bibitem{seiser_posch_1995}
\bibinfo{author}{C.~\surnamestart Seiser\surnameend},
  \bibinfo{author}{M.~\surnamestart Posch\surnameend},
  \bibinfo{author}{N.~\surnamestart Thompson\surnameend} \&
  \bibinfo{author}{L.~C. \surnamestart Kuhn\surnameend} (\bibinfo{year}{1995}):
  \emph{\bibinfo{title}{{E}ffect of transcription inhibitors on the
  iron-dependent degradation of transferrin receptor m{R}{N}{A}}}.
\newblock {\sl \bibinfo{journal}{J. Biol. Chem.}}
  \bibinfo{volume}{270}(\bibinfo{number}{49}), pp.
  \bibinfo{pages}{29400--29406}, \doi{10.1074/jbc.270.49.29400}.

\bibitemdeclare{article}{sharova_sharov_2009}
\bibitem{sharova_sharov_2009}
\bibinfo{author}{L.~V. \surnamestart Sharova\surnameend},
  \bibinfo{author}{A.~A. \surnamestart Sharov\surnameend},
  \bibinfo{author}{T.~\surnamestart Nedorezov\surnameend},
  \bibinfo{author}{Y.~\surnamestart Piao\surnameend},
  \bibinfo{author}{N.~\surnamestart Shaik\surnameend} \& \bibinfo{author}{M.~S.
  \surnamestart Ko\surnameend} (\bibinfo{year}{2009}):
  \emph{\bibinfo{title}{Database for m{R}{N}{A} half-life of 19 977 genes
  obtained by {D}{N}{A} microarray analysis of pluripotent and differentiating
  mouse embryonic stem cells}}.
\newblock {\sl \bibinfo{journal}{DNA Res.}}
  \bibinfo{volume}{16}(\bibinfo{number}{1}), pp. \bibinfo{pages}{45--58},
  \doi{10.1093/dnares/dsn030}.

\end{thebibliography}
\end{document}